# Roots of diversity relations

Peter Würtz[1] and Arto Annila[1,2,3,*]

[1]*Institute of Biotechnology*, [2]*Department of Physics* and [3]*Department of Biosciences, FI-00014 University of Helsinki, Finland*

**Abstract**

The species-area relationship is one of the central generalizations in ecology however its origin has remained a puzzle. Since ecosystems are understood as energy transduction systems, the regularities in species richness are considered to result from ubiquitous imperatives in energy transduction. From a thermodynamic point of view, organisms are transduction mechanisms that distribute an influx of energy down along the steepest gradients to the ecosystem's diverse repositories of chemical energy, *i.e.*, populations of species. Transduction machineries, *i.e.* ecosystems assembled from numerous species, may emerge and evolve toward high efficiency on large areas that hold more matter than small ones. This results in the well-known logistic-like relationship between the area and the number of species. The species-area relationship is understood, in terms of thermodynamics, to be the skewed cumulative curve of chemical energy distribution that is commonly known as the species-abundance relationship.

## 1. Introduction

Species–area relationships are frequently used to quantify, characterize and estimate diversity of biota [1,2,3,4,5]. Typically the number of species (*s*) in a taxon is shown versus the size of sampling area (*A*). For example, the number of bird species increases mostly monotonically with a decreasing slope on islands otherwise similar but increasingly larger in area [6]. The relationship is recognized as one of the few generalizations in ecology but its basis has remained obscure and hence also its functional form has been the subject of a long-standing debate [7,8].

Species richness data from many ecosystems over a wide range of areas follow the power law $s = cA^z$ where the slope *z* and intercept *c* are determined empirically from a log-log plot [9,10]. Nevertheless, this curve without an asymptote has been criticized as unphysical, *e.g.*, because the globe is finite [11,12]. Logistic models and sigmoidal curves are found to comply with observed species richness in large and bordered communities [13,14,15]. Moreover, the small island effect, *i.e.*, at the extreme of small sampling areas, the exponential form ($s \propto \log A$) [16] seems to account best for data [3,17,18,19].

Despite the nonconformity among the three species-area models, it has been pointed out that they could be approximations of a common but unknown functional form [20]. Such an anticipated universal relationship would indicate similarity in overall structural and functional organization of ecosystems rather than implying some common parameters for all ecosystems. In any case, the species richness depends on many other factors besides the area most notably insolation, temperature and rain fall. Species-energy theory [21] aims at taking these factors also into account.

Furthermore, it has been realized that the species-area relation is linked to species-abundance and distribution-abundance relations [22,23,24]. Abundant species make large fractions of the total number of individuals in an ecosystem, but curiously the probability density is skewed toward rarity in a log-normal like manner [5,25,26,27].

The species-area relationship could hardly be rationalized without making a connection to theory of evolution. Indeed, speciation as the source of diversity and its relation to the size of area became recognized already early on [28,29]. Evolutionary effects have continued to interest and call for understanding how non-equilibrium conditions affect the relationship [30] by contributing to an imbalance between extinction and colonization [31,32,33,34].

Thus, the puzzle about the origin of species-area relation appears particularly intricate because many factors affect the species richness although all of them seem to associate ultimately with energy, space and time. Thus we face the profound question, where do the roots of diversity–area relations stem from.

In this study the diversity relations are examined from the fundamental principle of increasing entropy that was recently formulated as an equation of motion [35]. The statistical physics formulation places the theory of evolution by natural selection [36] on the 2nd law of thermodynamics. According to the 2nd Law, flows of energy naturally select the steepest gradients. These are equivalent to the shortest paths by the principle of least action [37]. The thermodynamic formulation has been used to describe why natural distributions are skewed [38] and why standards such as chirality develop [39] as well as why genomes house diversity of non-expressed entities in addition to genes [40]. Also, the homeostatic nature of the global system, including its abiotic and biotic mechanisms, has been considered on the basis of imperatives in energy transduction [41]. These results are in agreement with earlier work based on the maximum entropy principle [42,43,44,45,46,47].

It is no new idea to consider the species-area relationship to stem from a general principle. The relationship has been understood by ecologists as a fundamental pattern of nature that extends far beyond and below the length scales of ecosystem organization [48,49]. The objective here is to clarify the fundamental reason why the number of species *vs.* area is described by the aforementioned functional forms, not to suggest a new species-area model. The description of an ecosystem as an energy transduction system is novel neither, but only until recently the thermodynamic formalism has been available to derive the regularities of ecosystem organization from the first principles.

## 2. Thermodynamic description of an ecosystem

Many spontaneous processes in nature, commonly referred to as *natural processes* [50], evolve toward more probable states by leveling differences in energy. The universal phenomenon of energy dispersal is also known by the principle of increasing entropy and by the 2nd law of thermodynamics. In accordance with classical texts [51,52,53,54], an ecosystem is regarded by thermodynamics as an open energy transduction network. Populations are diverse repositories of chemical energy and individual organisms are energy transformers that tap into available potentials to drain them. Flows of energy direct down along gradients when chemical reactions transform species from one repository to another. At the level of cells and organisms, the energy equalizing process is customarily referred to as metabolism. At the level of an ecosystem, the energy transforming structure is known as the food web.

The description of energy transduction by statistical physics remains at a formal level. All entities of an energy transduction system are described as energy densities [55]. In this way they can be compared with one and another to deduce which way energy will flow. In nature, potential energy differences among the entities, *e.g.*, populations of species are diminished by numerous processes that take place at molecular level, *e.g.* by photosynthesis, or at macroscopic level, *e.g.* by grazing.

An energy transduction network is thermodynamically self-similar in its structure at all levels of hierarchy. For example, atoms are the base constituents that make molecules. Likewise at a higher level of hierarchy, cells are the base constituents that make organisms that make populations. Owing to the scale independent-formalism, one may, at each and every level of hierarchical organization, transform the formal description to a model where entities are assigned with parameters and functions to account for their properties and mutual interactions.

The amount of chemical potential energy associated with a population of $N_j$ individuals is given by the chemical potential [56] $\mu_j = RT\ln[N_j\exp(G_j/RT)]$ where the Gibbs free energy $G_j$ is relative to the average energy $RT$. The concept of chemical potential is not restricted to molecules, but applies to all entities such as plants and animals that result from chemical reactions. A population of plant or animal species is associated with a chemical potential just as a population of molecular species. The chemical potential denotes essentially the trophic level height. In other words, the species at the top of food chain are thermodynamically 'expensive' to maintain by the long dissipative chain of energy transduction. The chemical potential is a valuable concept to deduce the structure of an ecosystem because the flows of energy equalize potentials. The stationary-state condition for chemical reactant populations [56] determines also plant and animal populations as results of numerous reactions.

In an ecosystem many reactions convert quanta $\Delta Q_{jk}$ of high-energy radiation from the Sun to chemical energy. Subsequently many additional reactions redistribute the resulting base potential among diverse repositories of chemical energy (Fig. 1). The overall energy transduction from the base production potential $\mu_1$ toward all other potentials $\mu_j$ takes the direction of increasing entropy $S$ [35]

$$S \approx \frac{1}{T}\sum_{j=1}^{J} N_j \left( \sum_{k=1}^{} \mu_k - \mu_j + \Delta Q_{jk} + RT \right) = R\sum_{j=1}^{J} N_j \left( \frac{A_j}{RT} + 1 \right). \quad (1)$$

The chemical potential difference, *i.e.*, the free energy, experienced by species $j$ is, in this context, usually referred to as affinity $A_j = \Sigma\mu_k + \Delta Q_{jk} - \mu_j$ or free energy relative to the average energy $RT$. The concept of $RT$ means that the system is sufficiently statistic [57], *i.e.*, a change in the energy influx is rapidly distributed within the entities of the system. Thus, no major potential differences will amount between the populations of species that interact with each other more frequently than the total energy content of evolving ecosystem changes. Nevertheless, a large variation in the energy influx due to the annual rhythm may drive huge population fluctuations. Also abrupt changes in conditions or mechanistic failures, *e.g.* due to a disease, may bring about a large imbalance.

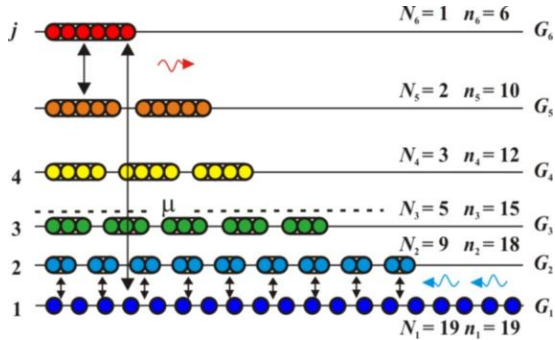

Fig. 1. Schematic distribution of chemical energy in a simple model ecosystem is described by an energy level diagram. The governing thermodynamic principle is exemplified by considering only one type of base constituents (atoms), but the result has been generalized for diverse base constituents [35]. The number of individuals at trophic level $j$ makes a population $N_j$. The corresponding density-in-energy $N_j\exp(G_j/RT)$ amounts from the number of base constituents $n_j = jN_j$ that are needed to assemble the population and from the invested energy $G_j$. For a species at a level $j$ in the food web many atoms and much energy are needed to propel its growth and to maintain it in the mature state. Species are equipped with mechanisms to generate these vital flows of energy by numerous reactions (arrows) that absorb high-energy or emit low-energy quanta (wavy arrows). Systems on larger areas, hence having access to more base constituents $N = \Sigma n_j$, will evolve to larger and more effective energy transduction machineries comprising more species. Coloring emphasizes that species differ from each other by their energy transduction properties, *i.e.*, phenotypes.

According to Eq. 1, the population $N_j$ may proliferate by acquiring ingredients $N_k$ and external energy $\Delta Q_{jk}$ from the surroundings, as long as $A_j > 0$. Likewise, when $A_j < 0$, the population $N_j$ is in for downsizing. When $A_j = 0$, the potential $\mu_j$ associated with $N_j$ of species $j$ matches the sum of potentials $\Sigma\mu_k$ of species $k$ and external energy $\Delta Q_{jk}$ that are vital for maintaining the population $N_j$. Finally, when all $A_j = 0$, the ecosystem has reached via numerous chemical reactions the maximum entropy state $S = R\Sigma N_j$, the stationary state of chemical non-equilibrium powered by solar flux. The species-area relationship, as will be shown below, is a consequence of the stationary-state structure of the ecosystem.

## 3. Distribution of chemical energy

During the course of evolution free energy is consumed and entropy increases at the rate [35]

$$\frac{dS}{dt} = \sum_{j=1}^{} \frac{dS}{dN_j}\frac{dN_j}{dt} = \frac{1}{T}\sum_{j=1}^{} v_j A_j \quad (2)$$

as the ecosystem moves to increasingly more probable states via numerous chemical reactions that adjust populations of species relative to one and other. During the evolutionary processes toward the thermodynamic steady state also new species may appear and old ones may disappear. New species will gain ground only when they are equipped with mechanisms that allow to them contribute to $S$. The old species will perish if their potentials are exhausted by others that have more efficient means of energy transformation.

To satisfy the balance equation, the population $N_j$ of species $j$ changes at the rate [35]

$$v_j = \frac{dN_j}{dt} = r_j \frac{A_j}{RT} \quad (3)$$

proportional to thermodynamic driving force $A_j$, *i.e.* a potential difference, by a mechanistic coefficient $r_j > 0$. The rate equation differs from phenomenological differential equations based on the law of mass-action that are used in population dynamics, *e.g.* for modeling colonization and extinction. The flow equation differs also from the logistic equation where a *constant* carrying capacity is taken proportional to the sampling area [51,58,59]. However, in reality there is no fixed carrying capacity but thermodynamic driving forces keep changing with changing populations that in turn affect the driving forces. In other words, the flows down along gradients keep changing due to the changing free energy landscape.

The interdependency among densities-in-energy means that when one species is consuming in its processes common resources, *e.g.*, base constituents the others have less. Even a small change in the initial conditions will affect the outcome later, hence by the definition [60] evolution is chaotic. For these reasons, it is in principle impossible to predict precisely trajectories of evolution and ensuing detailed structure of an ecosystem. Accordingly, there is no analytical form for the species-area relationship because it results from non-integrable and non-deterministic processes [35]. However, an effective approximation, in addition to the logistic and power law forms, is available.

## 4. Species-area relationship

Under a steady external flux of energy the ecosystem will eventually reach a stationary state, the climax corresponding to the maximum entropy. Then all thermodynamic driving forces have vanished and potentials across reactions are equal

$$dS/dt = 0 \Leftrightarrow \mu_j = \sum_k \mu_k + \Delta Q_{jk} \\ \Leftrightarrow N_j = \prod_{k=1} N_k \exp(-\Delta E_{jk}/RT) \quad (4)$$

where $\Delta E_{jk} = \Delta G_{jk} - \Delta Q_{jk}$. The condition of chemical non-equilibrium stationary state expresses the familiar pyramid of numbers by giving species in the order of increasing thermodynamic costs. The climax state corresponds to the thermodynamically most optimal populations $N_j$ at all trophic levels $j$. The non-equilibrium stationary state is maintained by incessant energy transduction powered by an external source. Such a system resides in the free energy minimum and will rapidly abolish any emerging energy differences. High through-flux is powering the climax state in agreement with the maximum power principle [61,62]. However, the stationary state does not have to house the maximum number of species that may have been encountered earlier during succession to the maturity. The succession culminates to the system of fewer species that are highly effective in energy transduction.

All potentials $\mu_j$ in the ecosystem ultimately tap into the base potential $\mu_1$, *i.e.*, couple to reactions that absorb solar energy (or extract from some other high-energy external source). Since the form given by Eq. 4 is difficult to analyze, we simplify the decreasing exponential partition (Eq. 4) by an average thermodynamic relation by expressing all interacting species $N_j$ in terms of stable (*i.e.* $G_1 = 0$) base constituents $N_1$, *i.e.*, atoms and external energy that is incorporated in the assembly processes. The average relation is merely a simplification of the energy transduction network (Eq. 4) but it allows us to depict the form of species-area relationship and compare the result with the relations that are known to account for the data.

The condition of thermodynamic stationary state

$$N_j = N_1^j \exp[(j-1)\Delta Q_1/RT] = \exp[\gamma(j-1)] \quad (5)$$

says how many stable base constituents $N_1$ and energy quanta $\Delta Q_1$ are required to maintain the population $N_j$ of species $j$ at the (trophic) level $G_j$, given concisely in units of the average base potential $\gamma = \ln N_1 + \Delta Q_1/RT$. The simplified stationary-state condition (Eq. 5) takes into account the larger number of base ingredients on larger areas but not that mechanisms of energy transduction evolve on larger areas more effective and efficient on larger areas than on small ones. Furthermore, the approximation that all species would have the same stoichiometric composition of base constituents $N_1$ on the average is reasonable for many biotic systems but it is not without exceptions. Therefore, parameters in the models of species-area relations are not universal as is apparent from many field studies.

The species-area relationship is essentially a consequence of conservation of matter. For a given influx of energy, the populations $N_j$ of all species $j$ (Eq. 5) each having the base constituents in numbers $n_j = jN_j$ (Fig. 1) are summed up to the total amount $N = \Sigma n_j$ that is taken proportional to the area $A$

$$N = \sum_{j=1}^s jN_j = \sum_{j=1}^s j\exp[\gamma(j-1)] = \alpha A. \quad (6)$$

When Eq. 6 is solved for the total number of species $s$ and plotted against increasing area $A$, the average thermodynamic relation gives understanding to the commonly used functional forms species-area curves (Fig. 2). However, it should be emphasized that Eq. 6 is not a model; it is the instructive approximation of Eq. 4 to deduce the structure of ecosystem's energy transduction network. The proportionality constant $\alpha$ consumes implicitly many factors. For example, the diverse base constituents originate mostly from the atmosphere above $A$, not from the ground that supplies nutrients. Therefore species-area relations are customarily extracted from samplings, ideally alike in constituents and energy input, differing only in their areas. Also different abiotic constituents, *e.g.*, water and carbon dioxide that couple to external energy, require different amounts of energy for activation. The many ingredients, in a form of base constituents and energy, influence how far the natural process may advance. They all are contained in Eqs. 1-3, but obviously it would be extremely challenging to model a large system in such a great detail.

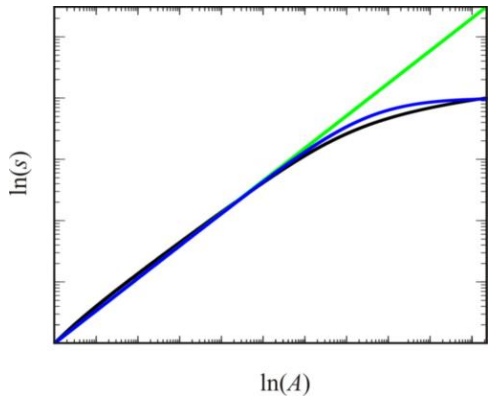

Fig. 2. Species (*s*) vs. area (*A*) relationship (black) is a cumulative curve of non-equilibrium stationary-state distribution of chemical energy in an ecosystem. The total amount of base constituents *N* in the system is taken proportional to the area *A*. The cumulative curve follows mostly the power law (green) but at large areas the logistic form (blue) accounts better for the statistical series. The units on axes depend on the energetics given by $\gamma$, units of measurements and proportionality constants.

The definition of species, implied by the index *j*, would mean that any two entities that can be distinguished from each other are distinct. In nature, entities distinguish from each other in interactions. Thus the definition of species is subject to the resolution that is available in the subjective detection process. The increment in index *j* is, therefore, not of primary interest when examining the functional form of species-area relationship.

The obtained form for the species-area curve (Fig. 2) is consistent with the data [5,12] and theoretical considerations [15,20]. At small areas it rises nearly exponentially, turns into the power-law form at larger areas and finishes in the logistic manner at the largest areas. The slope ln*s*/ln*A* diminishes with the increasing number of species. The correspondence to the power-law slope *z* is obtained from the derivative of *s*(*A*) and the relations to the parameters of logistic or exponential model by best fit of a particular data.

The debated question, does the species-area relationship have an asymptote, is not particularly meaningful because the thermodynamic objective is not to maximize the number of energy transformers of different kind but to arrive at the system in a stationary state with respect to its surroundings whatever number of species it takes. Thus, it is the surroundings that will ultimately dictate how high the system may possibly rise with its ingredients to make energy transformers. It is also emphasized that the sum over the species in Eq. 6 is open to the energy influx from the surroundings that is an ingredient along with the substances bound by Earth's gravitation.

## 5. Species-abundance relationship

To relate the species-area relationship with the species-abundance relationship, the sum over all species *j* in Eq. 6 is approximated by a convenient continuous function

$$\sum_{j=1}^{s} j \exp\left[\gamma(j-1)\right] \approx \int_{1}^{s} P(j)dj. \qquad (7)$$

The density function *P*(*j*) is the distribution of chemical energy. The skewed function peaks at the fractions that contribute most to entropy, *i.e.*, to energy dispersal and tails toward rare species' fractions (Fig. 3). The populations are in relation to their potentials. Those species that have mechanisms to tap into rich potentials on large areas are abundant, and they are also likely to find some resources on smaller areas to support a correspondingly smaller population. The thermodynamically expensive species consume large potentials hence they are rare even on large areas and unlikely to be found on smaller areas with in sufficient potentials.

According to the self-similar formulation of thermodynamics, also distributions of individuals are skewed, approximately log-normal, functions [38] in agreement with observations [5]. The most abundant bins of a distribution correspond to those individuals, *i.e.* mechanisms that contribute most to energy transduction. Likewise within a taxon, the density function *P*(*j*) *vs. j* displays a characteristic peak at the species richness that is identified to the intermediate size species [5]. It is these intermediate fractions that contribute the most to energy transduction. The variation of densities-in-energy among individuals in the same species is small in comparison with the total dispersal of energy in the entire ecosystem. This is to say that the individuals of the same species have approximately similar mechanisms of energy transduction whereas individuals of different species have distinctively different mechanisms. The skewed distributions have also been found in genomes [63] and rationalized using the 2$^{nd}$ Law [40]. The ubiquitous characteristics imply that the species-area and species-abundance relations are not only ecological relationships, but account for hierarchical organization of matter to dissipative systems in general.

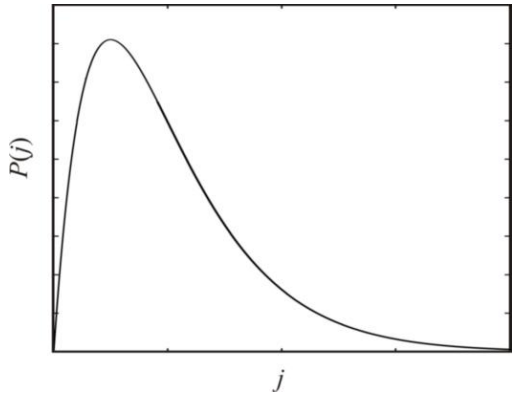

Fig. 3. Distribution of chemical energy among diverse chemical repositories $j$, *i.e.*, species according to Eq. 6. The probability density $P(j)$ of species-area curve is characteristically skewed toward rarity at high energy trophic levels $j$. The integral of $P(j)$ sums up all matter that is distributed among populations of all species $s$ in an ecosystem. When the total matter is taken proportional to the area $A$, the species-area relationship is obtained as the cumulative curve.

### 6. Species-area relation in evolution

At this point, it is insightful to describe effects of migration, speciation and extinction on the species-area relation using thermodynamics. Customarily the species-area relations are considered when there is a balance between immigration and *in situ* speciation and extinction. Obviously ecosystems evolve in space and time. Non-equilibrium conditions are expected to show in species-area relations.

According to the basic thermodynamic rationale, evolution as a whole is an energy transduction process. For any flow of energy, there is only one reason – an energy difference. Diverse differences in energy drive diverse flows that manifest, *e.g.*, as migration, speciation and extinction.

To begin with, the question, why there are so many species, calls for the answer. Functionalities of entities, *e.g.* organisms, appear in mutual interactions when they tap into various potentials by their phenotypic mechanisms. However, no single entity due to its finite composition may exhibit all possible functionalities to drain all conceivable sources of energy. This limits utilization of resources and promotes segregation of species for specialized and efficient functional roles to acquire chemical energy from specific sources. The populations of species themselves are repositories of energy for others to be consumed. Hence, diversity builds on diversity. In the quest to reduce all possible energy gradients, species evolve to thrive in ecological niches that are, thermodynamically speaking, basins in the free energy landscape. The diversification may also proceed within a species and manifest, e.g., as behavioral specialization, *i.e.* "division of labor".

The characteristic mechanisms of energy transduction are referred to as phenotypes that distinguish a species from another in the same system. According to the Lyapunov stability criterion that is given in terms of entropy [50,60], for any two species having nearly similar mechanisms, one will inevitably be excluded because such a system is unstable. The competitive exclusion principle is not limited to animals and plants but has been shown to account for the ubiquitous handedness of amino acids and nucleic acids as well [39].

The fitness criterion for natural selection, equivalent to the rate of entropy increase (Eq. 2), gives rise to increasingly economical and effective dissipative systems to consume various sources of free energy. Nevertheless, it may appear odd that species tend to evolve by retaining their ancestral ecological characteristics. From the thermodynamic viewpoint, an organism must sense an energy gradient for it to evolve. If there is not even a rudimentary or indirect mechanism available for a species to tap into a potential, the specific source of free energy provides no gradient for the species to direct its evolution. Hence, the particular species continues to diversify more readily along those gradients that are sensed by the mechanisms resulting from the ancestral development.

The phylogenetic conservatism may lead to an unusual species-area relation. When a species that is equipped with superior migratory mechanisms, such a bird species, happens to colonize a rich remote location, such as a large isolated island, phylogenetic conservatism may confine the ensuing diversification so that numerous mechanisms, *i.e.*, species will emerge however all with avian characteristics and none with truly optimal mechanisms for full terrestrial activity. Under those circumstances the number of species may become larger than expected on the basis of the islands area. Therefore, the ecosystem appears to be in a non-equilibrium state. To be more precise in wording, the ecosystem is stable, *i.e.* not subject to driving forces, but it is vulnerable to an eventual later colonization by more potent species from other ancestral lines that are more suited for terrestrial life. A single non-native species with superior mechanisms may rapidly drive numerous native species to extinction by consuming previously ineffectively and inefficiently used potentials. Obviously, a pioneering immigrant species that has specialized far away from its ancestral habitant and thus has given up its valuable virtues may fall as an easy prey for newly emerged predators.

It is also conceivable that a small remote location holds a lower number of species than expected on the basis of its area. Nowadays it is less likely that such an isolated and intact location could be found but certainly a newly surfaced volcanic island displays initially anomalously low species-area relation. When the area is small, all potentials are small and limited as well. Flows between the potentials are few and their rates are low. Also the rate of speciation is low and owing to the remote location, immigration rates are very low as well. It may then happen that the island lacks, *e.g.*, an entire genus. Then the ecosystem appears to be in a non-equilibrium state having too few species. More specifically, the state is stable until members of the 'missing' genus appear and expose the ecosystem to novel energy gradients. Then the diversification begins and brings up with time the number of species to the expected level.

The interdependent thermodynamic description takes into account effects that a new species introduces on all other species in an ecosystem. The new transduction mechanism puts the system in motion toward a new stationary state (Eq. 2). The species-area relations essentially states that for a new species ($s + 1$) to appear on increasingly larger areas, it will become increasingly more demanding, in thermodynamic terms, to meet the differentiation condition $dS/dN_{j+1} > \Sigma dS/dN_j$. For the new species to gain ground it must be able to increase entropy, *i.e.*, to disperse energy by its characteristic mechanisms more than could be achieved by increasing the populations of existing species.

A continent has more ingredients and more energy to fuel diverse flows that may combine so that a new species will emerge in comparison with a small island that is more likely to acquire new species by migration. An island next to the main land or a mountain top above a plain may acquire frequently new species. The small area may support some immigrants even below the aforementioned differentiation condition, but only for a limited time period. When the immigrants have over-depleted their vital potentials at the small location, they must leave to tap into potentials elsewhere or they will perish. Therefore, an adjacent island, just as a mountain top, that enjoys from a continuous influx of species may hold a larger number of species than would be expected only on the basis of its area. Such a state is usually referred to as a non-equilibrium state but when the influx is steady, the state is also steady.

## 7. Discussion

The thermodynamic description of an ecosystem as an energy transduction network and the view of species as energy transformers are not new ideas [51,52,53,54]. The new insight to biotic systems is provided by the 2$^{nd}$ Law of thermodynamics given as the equation of motion [35,37]. It reveals that the principle of increasing entropy and the theory of evolution by natural selection are in fact stating one and the same imperative; not describing opposing forces as it is often mistaken.

It is important to realize that the 2$^{nd}$ Law only states that differences in energy tend to diminish. Often it is one-sidedly thought that the 2$^{nd}$ Law would describe only the evolutionary course leading to diminishing densities-in-energy. This is the scenario at the cosmic scale. Here on Earth next to Sun, the imperative is the same but it is perceived differently. The flow of energy is also downward when the high-energy solar flux couples via chemical reactions to the low-energy matter on Earth. Consequently, chemical potential of matter is bound to increase when mechanisms that couple to the influx, happen to emerge.

The quest to diminish the energy difference with respect to the insolation directs evolution. Over the eons the machinery for the base production has emerged. The base production in turn, provides the high potential for other mechanisms to be consumed. In this way energy is distributed by diverse mechanisms downward to other repositories within the ecosystem and eventually dumped in as low-energy radiation in space. The imperative to level gradients increasingly more effectively and efficiently results in the characteristic regularities and relationships of nature. Intriguingly, such skewed distributions, *e.g.* of plants and animal populations, and sigmoid dispersion relations, *e.g.* species-area relation, are not only encountered in ecology but found also in many other contexts [64,65,38]. The thermodynamic formulation for the intricate and complex network of energy transduction of an ecosystem resembles the power-series derived from the concept of self-similarity [22] in accordance with the simplifying form of Eq. 6.

Despite the holistic view provided by thermodynamics, the self-consistent scale-independent description of energy transduction systems may appear abstract, especially as it seems to take no account on biological mechanisms, structures and functions. However, the entropy formula (Eq. 1) is deceptive in its conciseness. It describes energy densities in an entire ecosystem by every unit of matter $N_k$ and $N_j$ and by every quantum of energy $G_k$ and $G_j$, as well as by indexing all interactions by $j$ and $k$. Obviously it would require a detailed knowledge of all reactions, *e.g.*, the full atomic description of energy transduction, to establish the precise relationship between *s* and *A* for a particular ecosystem. Such a network of nested summations over all entities in Eq. 1 would be enormous and impractical, but the abridged form of Eq. 6 reveals the sigmoid diversity-area relation. It is, in terms of physics, a dispersion relation, *i.e.* the energy response function.

Properties of atoms, characteristics of molecules, functions of organisms, phenotypes of animals *etc.*, obtain their definitions in interactions. Also our observations are dissipative interactions [66] that classify individuals in diverse species. Increasingly powerful experimental methods allow us to distinguish finer and finer details. Consequently, the species is only a practical definition that refers to a particular class of densities-in-energy by emphasizing reactions of reproduction. Certainly, hereditary mechanisms are powerful, however irrespective of reproduction mechanisms the overall structure of any energy transduction is governed by the universal imperative to disperse energy down along gradients most rapidly.

Thermodynamic reasoning is simple. Systems, at all scales, evolve toward stationary states in their respective surroundings. Evolution is a natural process, a sequence of successive steps that makes no difference between inanimate and animate when devouring free energy. A small system will rapidly acquire mechanisms in succession, whereas for the global ecosystem it has taken eons to emerge via random variation with *de novo* mechanisms in the quest for a stationary state. For all systems it is the superior surrounding energy densities that command evolution. However, it takes mechanisms for energy to flow between the system and its surroundings. Intrinsic emergence of mechanisms or acquisition of them from the surroundings, unleash flows in the quest for the stationary state. However the equation of evolution, *i.e.*, the 2$^{nd}$ Law as the equation of motion cannot be solved because the flows affect the driving forces that in turn redirect the flows. Therefore, the courses of evolution are intricate and difficult to predict in detail.

For a long time there has been a search for the common ground to establish the many laws of ecology. The thermodynamics of open systems meets the early expectations of ecology as pronounced a century ago by Oscar Drude, an eminent plant ecologist. "Ecology has arisen from the need to unite originally separate branches of science in a new and natural doctrine; it is characterized by the breadth of its aims, and its peculiar power and strength in its ability to unite knowledge of the organic life with knowledge of its home, our earth. It assumes the solution of that most difficult as well as most fascinating problem which occupies the minds of philosophers and theologians alike, namely, the life history of the plants and animal worlds under the influences of space and time" [48].

**Acknowledgement**

We thank Erkki Annila and Sedeer El-Showk for enlightening discussions and valuable comments.